\documentclass{Interspeech2024}

\interspeechcameraready 
\usepackage{cite}
\usepackage{multirow, multicol}
\usepackage{makecell}
\usepackage{bbding}
\usepackage{enumitem}
\PassOptionsToPackage{hyphens}{url}
\usepackage{hyperref}
\usepackage{url}
\usepackage{authblk}
\title{Multi-Stage Speech Bandwidth Extension with Flexible Sampling Rate Control}
\name{Ye-Xin}{Lu}
\name{Yang}{Ai}
\name{Zheng-Yan}{Sheng}
\name{Zhen-Hua}{Ling}
\address{
  National Engineering Research Center of Speech and Language Information Processing, \\University of Science and Technology of China, Hefei, P. R. China}
\email{\{yxlu0102, zysheng\}@mail.ustc.edu.cn, \{yangai, zhling\}@ustc.edu.cn}


\begin{document}
\maketitle
\vspace{-1mm}
\begin{abstract}
\vspace{-1mm}
The majority of existing speech bandwidth extension (BWE) methods operate under the constraint of fixed source and target sampling rates, which limits their flexibility in practical applications.
In this paper, we propose a multi-stage speech BWE model named MS-BWE, which can handle a set of source and target sampling rate pairs and achieve flexible extensions of frequency bandwidth.
The proposed MS-BWE model comprises a cascade of BWE blocks, with each block featuring a dual-stream architecture to realize amplitude and phase extension, progressively painting the speech frequency bands stage by stage.
The teacher-forcing strategy is employed to mitigate the discrepancy between training and inference.
Experimental results demonstrate that our proposed MS-BWE is comparable to state-of-the-art speech BWE methods in speech quality.
Regarding generation efficiency, the one-stage generation of MS-BWE can achieve over one thousand times real-time on GPU and about sixty times on CPU.
% while achieving a fourfold acceleration for one-stage generation in inference efficiency.
\end{abstract}

\vspace{-1mm}
\begin{figure*}[ht!]
  \centering
  \includegraphics[width=\textwidth]{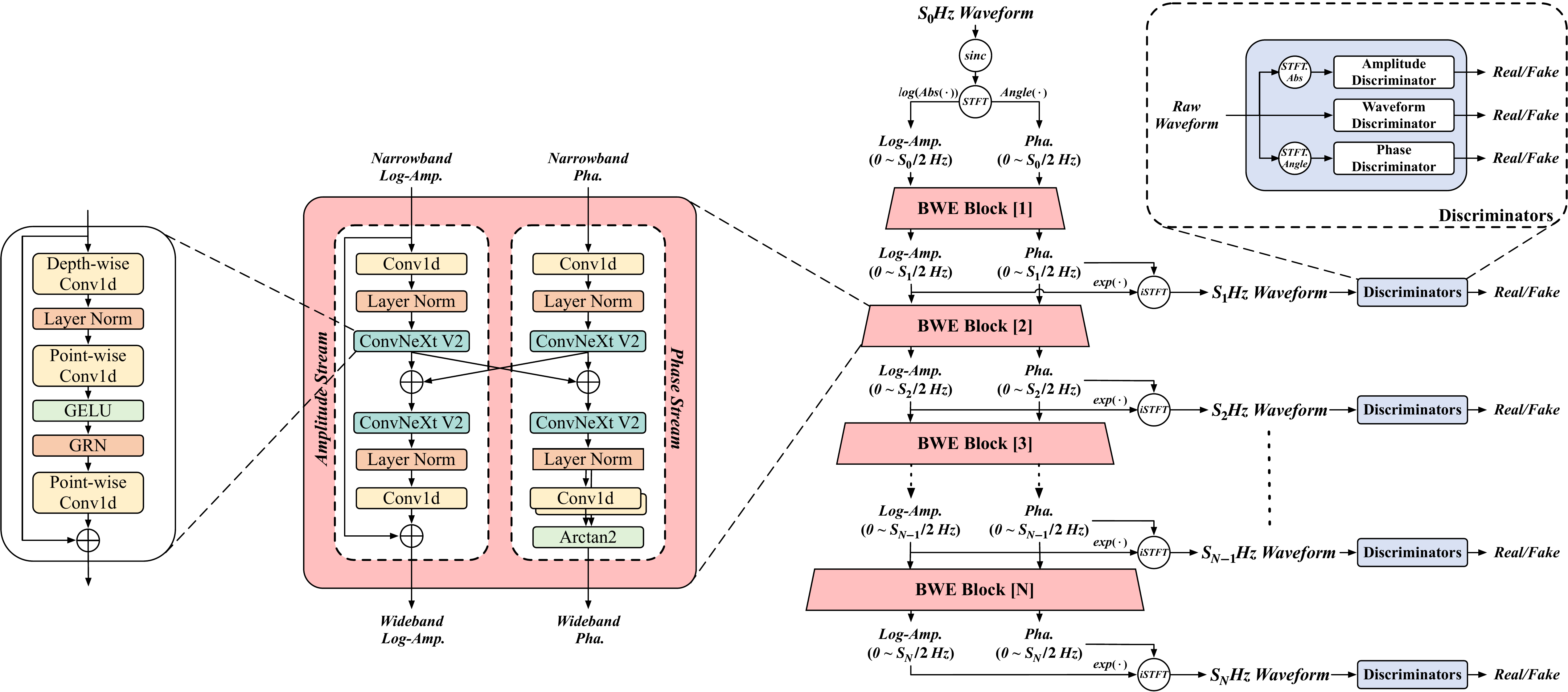}
  \caption{Overall structure of the proposed MS-BWE. The $sinc$ denotes the sinc filter interpolation, $\mathrm{Abs}(\cdot)$ and $\mathrm{Angle}(\cdot)$ denote the amplitude and phase calculation functions, $\log(\cdot)$ and $\exp(\cdot)$ denote the logarithmic and exponential functions, and ($0 \sim S_n/2$ Hz), $n\in\{0, 1, ..., N\}$ denotes that the effective frequency bands of the amplitude and phase spectra range from 0 to $S_n/2$ Hz.} 
  \label{fig: model}
\end{figure*}
\vspace{-1mm}
\section{Introduction}
\vspace{-1mm}
Speech bandwidth extension (BWE) aims to supplement the high-frequency components of narrowband speech signals, expanding the frequency bandwidth to enhance speech quality and intelligibility.
Traditional speech BWE methods utilized signal processing techniques to predict high-frequency residual signals and spectral envelopes, including source-filter-based methods \cite{makhoul1979high, chennoukh2001speech}, mapping-based methods \cite{carl1994bandwidth, unno2005robust, sadasivan2016joint}, and statistic methods \cite{pulakka2011speech, chen2004hmm, ohtani2014gmm, song2009study}.
However, these conventional methods suffered from bottlenecks in terms of model capabilities, resulting in the generation of over-smoothed spectral parameters \cite{ling2015deep}.
% Speech BWE can be seen as a constrained case of speech super-resolution (SR), which involves increasing the sampling rate of low-resolution speech signals in the time domain. 
% Low-resolution speech signals may contain aliasing artifacts of high-frequency components if they are obtained by subsampling. 
% In contrast, only low-frequency components are retained in the narrowband signals in the context of speech BWE.
% Consequently, Speech BWE presents more challenges compared to Speech SR.

With the development of deep learning, deep neural networks (DNNs) with powerful modeling capabilities were increasingly applied in the field of speech BWE.
DNN-based speech BWE methods can be broadly categorized into time-domain methods and frequency-domain methods.
In the category of time-domain methods, mapping-based methods learned the direct mapping from narrowband speech waveforms to their wideband counterparts \cite{kuleshov2017audio, ling2018waveform, birnbaum2019temporal, rakotonirina2021self}, while gradient-based methods leveraged diffusion models to recover wideband speech waveforms from the noised narrowband ones progressively \cite{lee2021nu, han2022nu, yu2023conditioning}.
In the frequency-domain category, vocoder-based methods adopted neural vocoders to recover the wideband speech waveforms from the extended mel-spectrograms \cite{liu2022neural}.
Spectrum-based methods chose to directly predict the wideband time-frequency transformed spectra \cite{mandel2023aero, shuai2023mdct, lu2024towards} from the narrowband ones and used inverse transformation to reconstruct the wideband speech waveforms.
While current speech BWE methods have exhibited promising performance, they were constrained by predefined source and target sampling rates. 
Due to the varying sampling rates of speech in different practical application scenarios, establishing a separate model for each sampling rate pair would result in a significant memory footprint requirement and a lack of flexibility in model utilization.
% Although some of them can simultaneously handle inputs with varying sampling rates using a unified model \cite{liu2022neural, han2022nu, yu2023conditioning, lu2024towards}, the target sampling rate remained fixed, resulting in inefficient utilization of model parameters and limiting their flexibility in practical applications.

To this end, we propose MS-BWE, a multi-stage speech BWE model that can handle flexible source and target sampling rate pairs.
The proposed MS-BWE comprises a series of BWE blocks, facilitating extensions among a set of sampling rates from low to high.
For each BWE block, we follow our previous work \cite{lu2024towards} to use a dual-stream architecture to predict the wideband log-amplitude and phase spectra from the narrowband counterparts derived by short-time Fourier transform (STFT).
The extended speech waveforms of each BWE block can be obtained from the extended log-amplitude and phase spectra using inverse STFT (iSTFT).
Furthermore, to mitigate the discrepancy between training and inference as well as achieve flexible extension across multiple stages of sampling rates, we employ the teacher-forcing strategy \cite{bengio2015scheduled} to randomly use either the real log-amplitude and phase spectra or the generated ones from the prior BWE block as the input of current BWE block with scheduled sampling.
Experimental results demonstrate that our proposed MS-BWE is comparable to the state-of-the-art (SOTA) speech BWE methods in speech quality across flexible sampling rates with a unified model.
In terms of generation efficiency, the one-stage generation of our proposed MS-BWE can generate 48 kHz waveform samples 1271.81 times faster than real-time on a single NVIDIA RTX 4090 GPU and 59.70 times faster than real-time on a single CPU.
% Compared to the SOTA speech BWE methods, we can also achieve at least a fourfold acceleration on both GPU and CPU.
\vspace{-1mm}
\section{Methodology}
\vspace{-1mm}
\subsection{Overview}
\vspace{-1mm}
The overall structure of the proposed MS-BWE is illustrated in Fig.~\ref{fig: model}.
Given a set of sampling rates $\mathbb{S}=\{S_0, S_1, ..., S_N\}$ Hz, where $S_0 < S_1 < ... <S_N$, the proposed MS-BWE aims to realize flexible extension between any source and target sampling rate pair $(S_i, S_j)$, $0 \leq i < j \leq N$.
To achieve this, the proposed MS-BWE comprises $N$ BWE blocks, with each block sequentially implementing the extension between adjacent sampling rates.
Following our previously proposed AP-BWE \cite{lu2024towards}, we design each BWE block to extend the speech waveforms at the spectral level through parallel amplitude and phase streams.

During the training stage, the narrowband waveform with a sampling rate of $S_0$ Hz is first interpolated to a narrowband waveform with the sampling rate of $S_N$ using a sinc filter. 
Subsequently, the corresponding narrowband amplitude and phase spectra are extracted from the sinc-interpolated narrowband waveform through STFT and then fed to the first BWE block.
For the $n$-th BWE block, where $ 1 < n \leq N$, it randomly samples either the real log-amplitude and phase spectra extracted from the sinc-interpolated $S_{n-1}$ Hz waveform or the generated ones from the previous BWE block as inputs.
At the inference stage, for the speech BWE from a source sampling rate of $S_i$ Hz to a target sampling rate of $S_j$ Hz, the narrowband $S_i$ Hz waveform first undergoes sinc interpolation and then STFT to extract the narrowband log-amplitude and phase spectra, which are then fed into the $(i+1)$-th BWE block. 
After undergoing $(j-i)$ stages of generation, the $j$-th BWE block outputs the extended log-amplitude and phase spectra with the effective bandwidth of $S_j/2$ Hz, which are further transformed back into a $S_j$ Hz waveform via iSTFT.
The details of the model structure and training criteria are described as follows.
\vspace{-1mm}
\subsection{Model structure}
\vspace{-1mm}
As depicted in Fig~\ref{fig: model}, the proposed MS-BWE comprises a cascade of BWE blocks, each taking a pair of narrowband log-amplitude spectrum and phase spectrum as inputs and generating the corresponding extended spectra through parallel amplitude and phase streams.
Both the amplitude stream and the phase stream utilize the ConvNeXt V2 \cite{woo2023convnext} as their foundational backbone.
Similar to the ConvNeXt \cite{liu2022convnet} used in AP-BWE \cite{lu2024towards}, the ConvNeXt V2 features a depth-wise convolutional layer and a pair of point-wise convolutional layers, interleaved with layer normalization \cite{ba2016layer} and Gaussian error linear unit (GELU) activation \cite{hendrycks2017gaussian}.
Differently, ConvNeXt V2 added a new global response normalization (GRN) layer after the GELU activation to enhance inter-channel feature competition.
% In our preliminary experiments, we observed that ConvNeXt V2 exhibited stronger modeling capabilities compared to the original ConvNeXt in the speech BWE task.

On the basis of ConvNeXt V2, the amplitude stream and phase stream share a similar architecture.
Each stream consists of two ConvNeXt V2 blocks and employs two convolutional layers with layer normalization on both sides for feature-dimensionality expansion and restoration, respectively.
For the amplitude stream, the output convolutional layer predicts the residual log-amplitude spectrum, which is subsequently added to the input narrowband one to obtain the extended log-amplitude spectrum.
Nevertheless, due to the phase-wrapping issue, the phase stream utilizes the parallel wrapped phase estimation architecture \cite{ai2023neural} to predict the extended phase spectrum directly, which comprises two parallel convolutional layers to output the pseudo-real part and imaginary part components and further calculate the wrapped phase spectrum with the activation of the two-argument arc-tangent (Arctan2) function.
\vspace{-1mm}
 % (i.e., $n=1, 2, 3, 4$ for 24 kHz $\rightarrow$ 48 kHz, 16 kHz $\rightarrow$ 48 kHz, 12 kHz $\rightarrow$ 48 kHz, and 8 kHz $\rightarrow$ 48 kHz, respectively)
\begin{table*}[htbp!]
  \caption{Experimental results for BWE methods evaluated on the VCTK-0.92 dataset with the target sampling rate of 48 kHz, where in RTF ($a\times$) represents $a$ times real-time and $n$ represents the $n$-stage generation. The \textbf{bold} and \underline{underlined} numbers indicate the optimal and sub-optimal results, respectively.}
  \label{tab: results_48k}
  \centering
  \resizebox{\textwidth}{!}{
  \begin{tabular}{l|cc|cc|cc|cc|cc}
    \toprule
    \multirow{2}{*}{Method} & \multicolumn{2}{c|}{8 kHz $\rightarrow$ 48 kHz} & \multicolumn{2}{c|}{12 kHz $\rightarrow$ 48 kHz} & \multicolumn{2}{c|}{16 kHz $\rightarrow$ 48 kHz} & \multicolumn{2}{c|}{24 kHz $\rightarrow$ 48 kHz} & \multirow{2}{*}{RTF (CPU)} & \multirow{2}{*}{RTF (GPU)} \\
    \cmidrule{2-9}
    & LSD & ViSQOL & LSD & ViSQOL & LSD & ViSQOL & LSD & ViSQOL &  &  \\
    \midrule
    sinc & 2.94 & 2.07 & 2.75 & 2.09 & 2.57 & 2.26 & 2.17 & 2.99 & - & - \\
    \midrule
    NU-Wave2 \cite{han2022nu} & 1.09 & 2.48 & 0.94 & 2.75 & 0.86 & 3.00 & 0.72 & 3.74 & 92.5836 (0.01$\times$) & 0.5195 (1.92$\times$) \\
    UDM+ \cite{yu2023conditioning} & 1.03 & 2.81 & 0.88 & 3.08 & 0.79 & 3.35 & 0.64 & 4.02 & 74.0332 (0.01$\times$) & 0.8335 (1.20$\times$)\\
    mdctGAN \cite{shuai2023mdct} & 0.93 & 2.95 & 0.90 & 2.96 & 0.82 & 3.15 & 0.72 & 3.58 & 0.2461 (4.06$\times$) & 0.0129 (77.80$\times$)\\
    AP-BWE\cite{lu2024towards} & \textbf{0.85} & \textbf{3.32} & \textbf{0.79} & \textbf{3.46} & \textbf{0.72} & \underline{3.63} & \textbf{0.62} & \textbf{4.17} & 0.0551 (18.14$\times$) & 0.0034 (292.28$\times$) \\
    \midrule
    \multirow{2}{*}{MS-BWE} & \multirow{2}{*}{\textbf{0.85}} & \multirow{2}{*}{\underline{3.31}} & \multirow{2}{*}{\textbf{0.79}} & \multirow{2}{*}{\underline{3.44}} & \multirow{2}{*}{\underline{0.73}} & \multirow{2}{*}{\textbf{3.65}} & \multirow{2}{*}{\underline{0.63}} & \multirow{2}{*}{\underline{4.14}} & 0.0167 * $n$ & \textbf{0.0008 * $\boldsymbol{n}$} \\
    & & & & & & & & & (59.70 / $n$ $\times$) & \textbf{(1271.81 / $\boldsymbol{n}$ $\times$)} \\
    \bottomrule
  \end{tabular}}
\end{table*}
\vspace{-1mm}
\subsection{Training criteria}
\vspace{-1mm}
\subsubsection{Loss functions}
\vspace{-1mm}
To avoid the generation of over-smoothed spectra, we employ the training approach of the generative-adversarial network (GAN) to define $N$ sets of discriminators on the extended speech waveforms of the $N$ BWE blocks.
Within each set of the discriminators, we first define the waveform discriminator at the waveform level, which derives from the sub-discriminator of the multi-scale discriminator \cite{kumar2019melgan, kong2020hifi}.
Furthermore, we respectively define the amplitude discriminator and phase discriminator to enhance the realism of extended amplitude and phase spectra, which are borrowed from the sub-discriminators of the multi-resolution amplitude and phase discriminators \cite{lu2024towards}.

We use the hinge GAN loss \cite{zeghidour2021soundstream} to jointly train the MS-BWE model and the discriminators.
For the generator loss, besides the GAN losses, we follow AP-BWE \cite{lu2024towards} to define spectral losses on the outputs of each BWE block, including log-amplitude mean square error (MSE) loss, phase anti-wrapping losses \cite{ai2023neural}, and short-time complex spectral MSE loss.
\vspace{-1mm}
\subsubsection{Teacher-forcing strategy}
\vspace{-1mm}
\label{sec: teacher-forcing}
Since the speech BWE process is carried out stage by stage, it can lead to two types of mismatch between training and inference: 1) In the training stage, each BWE block only samples the real amplitude and phase as inputs, i.e., the $N$ BWE blocks are trained separately. For the inference from $S_i$ Hz to $S_j$ Hz, if $j-i >1$, the $i+2, i+3, ..., j$-th BWE block need to receive the outputs of the previous block as inputs, which mismatches with the real inputs in training.
2) In the training stage, the $n$-th BWE block only samples the generated amplitude and phase from the $(n-1)$-th block as inputs, where $1<n\leq N$.
For the inference from $S_i$ Hz to $S_j$ Hz, if $j > i > 1$, the $(i+1)$-th BWE block needs to receive the real amplitude and phase as inputs, which mismatches with the generated inputs in training.

% To achieve flexible bandwidth extension, a straightforward approach is to train the $N$ BWE blocks separately. 
% However, this can lead to the first type of mismatch between training and inference, where the inputs of each BWE block during training are real amplitude and phase, while during multi-stage inference, the intermediate BWE blocks need to receive the outputs of the previous block as inputs.
% However, if the model only samples its self-generated amplitude and phase spectra as inputs, the second type of mismatch would occur when inferring from the intermediate BWE blocks with real amplitude and phase as inputs.

In order to address these two types of mismatch issues, we propose to employ the teacher-forcing strategy with schedule sampling \cite{bengio2015scheduled}.
During the training stage, except for the first BWE block, the remaining blocks randomly sample their inputs from either the real amplitude and phase or the extended ones from the previous block with a ratio.
This teacher-forcing ratio is scheduled to decrease progressively every mini-batch, transitioning the model from relying more on real samples to using more generated samples.
This strategic shift aids the model in better adapting to its own generated outputs, consequently mitigating the discrepancy between training and inference.
\vspace{-1mm}
\section{Experiments}
\vspace{-1mm}
\subsection{Datasets and experimental setup}
\vspace{-1mm}
We used the publicly available VCTK-0.92 dataset \cite{yamagishi2019cstr} for our experiments, which contains about 44 hours of speech recording at a sampling rate of 48 kHz from 110 speakers. 
Following the same configuration of previous works \cite{han2022nu, yu2023conditioning}, we split the data into training and test sets.
The experiments were performed with commonly used sampling rates, where $\mathbb{S}=\{8000, 12000, 16000, 24000, 48000\}$ and consequently $N=4$. 
To obtain narrowband speech signals, we downsampled the 48 kHz speech waveforms with the sinc filter to decimate the high-frequency components without any alias.

For training the MS-BWE model, all the speech recordings were sliced into 8000-sample-point clips and subsequently processed by STFT to extract amplitude and phase spectrum with the FFT point number, Hanning window size, and hop size of 1024, 320, and 80, respectively.
The teacher-forcing ratio was set initially to 0.75 and scheduled to decay with a factor of 0.999995 every mini-batch, where the batch size was set to 16.
The MS-BWE model was trained until 500k steps using the Adam optimizer \cite{loshchilov2017decoupled} with $\beta_1=0.8$, $\beta_2=0.99$, and weight decay $\lambda=0.01$.
The learning rate was set initially to $2\times 10^{-4}$ and scheduled to decay with a factor of 0.999 every epoch. \footnote{Audio samples of the proposed MS-BWE can be accessed at \href{https://yxlu-0102.github.io/MS-BWE-demo}{https://yxlu-0102.github.io/MS-BWE-demo}.}

\vspace{-1mm}
\subsection{Baseline methods}
\vspace{-1mm}
We first used the sinc filter interpolation as the lower-bound baseline, and compared our proposed MS-BWE with two diffusion-based methods (NU-Wave 2 \cite{han2022nu} and UDM+ \cite{yu2023conditioning}), a modified discrete cosine transform (MDCT) spectrum-based method (mdctGAN \cite{shuai2023mdct}), and an amplitude-phase spectrum-based method (AP-BWE \cite{lu2024towards}).
For the common experiments targeting 48 kHz, we first used the reproduced NU-Wave 2 checkpoint and the official UDM+ checkpoint in UDM+'s official implementation\footnote{\href{https://github.com/yoyololicon/diffwave-sr}{https://github.com/yoyololicon/diffwave-sr}.}.
We further adopted the official checkpoints of mdctGAN and the unified AP-BWE in their open-source implementations\footnote{\href{https://github.com/neoncloud/mdctGAN}{https://github.com/neoncloud/mdctGAN}.}\footnote{\href{https://github.com/yxlu-0102/AP-BWE}{https://github.com/yxlu-0102/AP-BWE}.}, and re-trained them for experiments with other pairs of source and target sampling rates.

\vspace{-1mm}
\subsection{Evaluation metrics}
\vspace{-1mm}
For extended speech quality evaluation, we first utilized the commonly used log-spectral distance (LSD) as the objective evaluation metric.
% with the FFT-point numbers, Hanning window size, and hop size set to 2048, 2048, and 512 for spectral extraction.
Additionally, we employed the virtual speech quality objective listener (ViSQOL) \cite{chinen2020visqol} to evaluate the overall speech quality, which ranges from 1 to 4.75 at 16 kHz and from 1 to 5 at 48 kHz.
For speech signals extended to 12 kHz or 24 kHz, we respectively resampled them to 16 kHz and 48 kHz to evaluate the ViSQOL score.
For LSD, lower values indicate better performance, while for ViSQOL, the higher, the better.
To assess generation efficiency, we used the real-time factor (RTF), which is defined as the ratio between the total inference time on the test set and the total duration of the test set.
In our implementation, we calculated the RTFs on a single RTX 4090 GPU and a single Intel(R) Xeon(R) Silver 4310 CPU (2.10 GHz).
\vspace{-1mm}
\begin{table*}[htbp!]
  \caption{Experimental results for BWE methods evaluated on the VCTK-0.92 dataset with flexible source and target sampling rates, where $a$M * $b$ indicates $b$ models with $a$M parameters each are required for all the speech BWE implementations.}
  \label{tab: results_flexible}
  \centering
  \resizebox{\textwidth}{!}{
  \begin{tabular}{l|cc|cc|cc|cc|cc|cc|c}
    \toprule
    \multirow{2}{*}{Method} & \multicolumn{2}{c|}{8 kHz $\rightarrow$ 12 kHz} & \multicolumn{2}{c|}{8 kHz $\rightarrow$ 16 kHz} & \multicolumn{2}{c|}{8 kHz $\rightarrow$ 24 kHz} & \multicolumn{2}{c|}{12 kHz $\rightarrow$ 16 kHz} & \multicolumn{2}{c|}{12 kHz $\rightarrow$ 24 kHz} & \multicolumn{2}{c|}{16 kHz $\rightarrow$ 24 kHz} & \multirow{2}{*}{\# Param.} \\
    \cmidrule{2-13}
    & LSD & ViSQOL & LSD & ViSQOL & LSD & ViSQOL & LSD & ViSQOL & LSD & ViSQOL & LSD & ViSQOL & \\
    \midrule
    mdctGAN & 0.64 & 4.72 & 0.78 & 4.57 & 0.89 & 3.79 & \textbf{0.55} & 4.81 & 0.75 & 3.86 & 0.66 & 4.03 & 103M * 6 \\
        AP-BWE & \textbf{0.62} & \textbf{4.81} & \underline{0.73} & \textbf{4.69} & \underline{0.83} & \textbf{3.87} & \textbf{0.55} & \textbf{4.89} & \textbf{0.72} & \textbf{3.96} & \underline{0.61} & \textbf{4.11} & \underline{30M * 3} \\
    \midrule
    MS-BWE & \textbf{0.62} & \underline{4.74} & \textbf{0.72} & \underline{4.59} & \textbf{0.82} & \underline{3.81} & \underline{0.56} & \underline{4.82} & \underline{0.74} & \underline{3.90} & \textbf{0.59} & \underline{4.08} & \textbf{43M * 1} \\
    \bottomrule
  \end{tabular}}
\end{table*}
\vspace{-1mm}
\section{Results and Analysis}
\vspace{-1mm}
\subsection{Comparison with baseline methods}
\vspace{-1mm}
\subsubsection{Many-to-one speech BWE}
\vspace{-1mm}
The current mainstream methods \cite{han2022nu, yu2023conditioning, shuai2023mdct, lu2024towards} all performed speech BWE in a many-to-one manner with multiple source sampling rates and a fixed target sampling rate.
Therefore, we first compared our proposed MS-BWE with these SOTA baseline methods in this configuration.
For source sampling rates of 8 kHz, 12 kHz, 16 kHz, and 24 kHz, our proposed MS-BWE respectively utilized 4, 3, 2, and 1 BWE block(s) for generations, while NU-Wave 2, UDM+, and AP-BWE utilized unified models and mdctGAN used four separate models.

As shown in Table~\ref{tab: results_48k}, our proposed MS-BWE achieved comparable performance with AP-BWE and far surpassed other baseline methods in terms of extended speech quality.
Compared to the sub-optimal mdctGAN, the MS-BWE exhibited significant improvements of 8.6\%, 12.2\%, 10.9\%, and 12.5\% in LSD as well as 12.2\%, 16.2\%, 15.8\%, and 15.6\% in ViSQOL for source sampling rates of 8 kHz, 12 kHz, and 16 kHz, and 24 kHz, respectively.
Compared to AP-BWE, the performance of MS-BWE was slightly inferior, which may be attributed to the simultaneous optimizations of multiple training objectives in our model.
In terms of generation efficiency, although AP-BWE has achieved at least four times faster than other baselines, it still employed a unified model for all source sampling rates, resulting in the wastage of model parameters.
For our proposed MS-BWE, the efficiency of one-stage generation (e.g., 24 kHz $\rightarrow$ 48 kHz) can reach an approximately fourfold acceleration than that of AP-BWE on both GPU and CPU. 
This notable enhancement was attributed to the fact that, for the one-stage generation, MS-BWE saved nearly $3/4$ of the parameters compared to AP-BWE, achieving a more efficient parameter utilization.
Similarly, the $n$-stage generation of our proposed MS-BWE can generate 48 kHz waveform samples about $1271.81 / n$ times faster than real-time on a single GPU and about $59.70 / n$ times faster than real-time on a single CPU. 

\vspace{-1mm}
\subsubsection{Many-to-many speech BWE}
\vspace{-1mm}
We further compared our proposed MS-BWE with two optimal baseline methods (i.e., mdctGAN and AP-BWE) in a many-to-many manner of speech BWE to evaluate our method's flexibility.
As shown in Table~\ref{tab: results_flexible}, our proposed MS-BWE apparently outperformed mdctGAN, and performed comparably to AP-BWE in LSD but lagged behind in ViSQOL, which was consistent with the results in Table~\ref{tab: results_48k}.
It is noteworthy that, for the sampling rate set $\mathbb{S}=\{S_0, S_1, ..., S_N\}$, the approaches which can only handle one pair of source and target sampling rates at a time (e.g., mdctGAN), required $\frac{N^2+N}{2}$ individual models to achieve all the extensions between these sampling rates.
Even for methods that can simultaneously handle multiple source sampling rates (e.g., AP-BWE), at least $N$ independent models were still required.
Nevertheless, our proposed MS-BWE demonstrated the ability to realize flexible extensions across the sampling rate set using a unified model, while maintaining comparable performance to these individual baseline models.
Therefore, as shown in Table 2, the total parameter requirement for all speech BWE implementations of our model was significantly lower than the other two baseline models, allowing it to be flexibly applied to resource-constrained scenarios.

\vspace{-1mm}
\subsection{Analysis on training strategies}
\vspace{-1mm}
To verify the effectiveness of the teacher-forcing strategy on the overall model performance, we conducted analysis experiments by using different training strategies, and the experimental results are presented in Table~\ref{tab: results_analysis}.
Initially, we trained the MS-BWE model without sampling from its own generated intermediate results (denoted as ``Never Sampling'').
In this scenario, each BWE block was trained separately using real narrowband and wideband spectra pairs.
The results indicated that this training strategy yielded optimal performance for one-stage generation (e.g., 24 kHz $\rightarrow$ 48 kHz), while the prediction errors accumulated as the number of stages increased, corresponding to the first type of mismatch discussed in Section~\ref{sec: teacher-forcing}.
% which was reasonable as training separately would make each BWE block focus more on the extension of its predefined source-target sampling rate pair.
Subsequently, we trained the model to always sample from its own generated spectra (denoted as ``Always Sampling''). 
In contrast to the ``Never Sampling'' scenario, the experimental results demonstrated that the model performed well for the whole $N$-stage generation task (i.e., 8 kHz $\rightarrow$ 48 kHz), while the performance gradually collapsed as the number of the generation stages decreased, aligning with the second type of mismatch mentioned in Section~\ref{sec: teacher-forcing}.
Ultimately, with the implementation of the teacher-forcing strategy with scheduled sampling (denoted as ``Scheduled Sampling''), the model achieved a trade-off in multi-stage generation tasks by sampling from either the generated or real amplitude and phase spectra with a scheduled ratio. 
While it may not achieve optimal performance in each one-stage task, this approach resulted in flexible and relatively high-quality extension across the sampling rate set.
\vspace{-1mm}
\begin{table}[t!]\Huge
  \caption{Experimental results for the analysis of different training strategies with the target sampling rate of 48 kHz.}
  \label{tab: results_analysis}
  \centering
  \resizebox{\linewidth}{!}{
  \begin{tabular}{l|cc|cc|cc}
    \toprule
    \multirow{2}{*}{Training Strategy} & \multicolumn{2}{c|}{8kHz$\rightarrow$48kHz} & \multicolumn{2}{c|}{16kHz$\rightarrow$48kHz} & \multicolumn{2}{c}{24kHz$\rightarrow$48kHz} \\
    \cmidrule{2-7}
    & LSD & ViSQOL & LSD & ViSQOL & LSD & ViSQOL \\
    \midrule
    Never Sampling & 0.97 & 2.95 & 0.75 & 3.50 & \textbf{0.61} & \textbf{4.23} \\
    Always Sampling & 0.87 & 3.20 & 0.82 & 3.47 & 0.77 & 3.77 \\
    Scheduled Sampling & \textbf{0.85} & \textbf{3.31} & \textbf{0.73} & \textbf{3.65} & 0.63 & 4.14 \\
    \bottomrule
  \end{tabular}}
\end{table}
\vspace{-1mm}
\section{Conclusions}
\vspace{-1mm}
In this paper, we proposed MS-BWE, a multi-stage speech BWE model that achieves flexible extensions of a sampling rate set from low to high.
The proposed MS-BWE model was GAN-based, the generator of which consisted of multiple BWE blocks, each comprising parallel amplitude and phase streams to explicitly predict the high-frequency amplitude and phase components from the narrowband spectra.
Discriminators were employed on the outputs of each BWE block to further enhance the realism of each extended speech waveform at both the waveform level and the spectral level.
To mitigate the discrepancy between training and inference, we employed the teacher-forcing strategy to randomly introduce the real amplitude and phase to the intermediate BWE blocks.
Experimental results demonstrated that our proposed MS-BWE performed comparably to the SOTA baseline methods in flexible speech BWE tasks with a unified model while ensuring remarkable efficiency.
In summary, through the stage-wise speech BWE process, our method made full use of the model parameters and demonstrated the potential for practical applications in resource-constrained scenarios.
% Applying it to a broader range of sampling rate control will be the focus of our future work.

% our approach enhanced generation efficiency and flexibility while ensuring the quality of extended speech.
% \section{Discussion}

% Authors must proofread their PDF file prior to submission, to ensure it is correct. Do not rely on proofreading the \LaTeX\xspace source or Word document. \textbf{Please proofread the PDF file before it is submitted.}

% \section{Acknowledgements}
% Acknowledgement should only be included in the camera-ready version, not in the version submitted for review.
% The 5th page is reserved exclusively for \red{acknowledgements} and  references. No other content must appear on the 5th page. Appendices, if any, must be within the first 4 pages. The acknowledgments and references may start on an earlier page, if there is space.

\bibliographystyle{IEEEtran}
\bibliography{mybib}

\end{document}